\documentclass[aps,prl,twocolumn,preprintnumbers,showpacs,amsmath,amssymb]{revtex4}

\usepackage[usenames,dvipsnames]{color}
\usepackage[usenames]{color}
\usepackage{graphicx}
\usepackage{subfigure}

\begin{document}
\title{Nondestructive Detection of Polar Molecules via Rydberg Atoms}
\author{M. Zeppenfeld}
\email{martin.zeppenfeld@mpq.mpg.de}
\affiliation{Max-Planck-Institut f\"ur Quantenoptik, Hans-Kopfermann-Stra{\ss}e~1, 85748 Garching, Germany}


\begin{abstract}
A highly sensitive, general, and preferably nondestructive technique to detect polar molecules would greatly advance a number of fields, in particular quantum science with cold and ultracold molecules. Here, we propose using resonant energy transfer between molecules and Rydberg atoms to detect molecules. Based on an energy transfer cross section of $>10^{-6}$\,cm$^2$ for sufficiently low collision energies, a near unit efficiency non-destructive detection of basically any polar molecule species in a well defined internal state should be possible.
\end{abstract}

\pacs{34.90.+q}

\keywords{Cold molecules, Rydberg atoms}

\maketitle

The ability to detect single particles, e.g.\ single electrons or single photons, has been fundamental to the success of quantum science. Thus, highly efficient superconducting single-photon detectors have been essential for the recent simultaneous elimination of multiple loop holes for measuring violations of Bell inequalities~\cite{Shalm15,Giustina15}, and the ability to detect single atoms with single site resolution in optical lattices~\cite{Bakr09} has allowed investigation of quantum many-body physics at a single site level~\cite{Endres11,Cheneau12}.

A relatively new subfield of quantum science is the investigation of cold and ultracold polar molecules. Motivated by applications such as quantum information processing~\cite{DeMille02}, precision measurements~\cite{DeMille08}, or investigation of dipolar quantum gases~\cite{Baranov12}, substantial progress has recently been made in preparing controlled molecular ensembles~\cite{Gloeckner15b,Prehn16,Norrgard16,Moses16}.

Due to their complex structure, detecting polar molecules can be a formidable challenge. In the context of cold polar molecules, a number of detection techniques have been used. These include resonantly enhanced multi photon ionization (REMPI)~\cite{Bethlem00,Bertsche10}, light induced fluorescence (LIF)~\cite{Weinstein98,Shuman09}, absorption spectroscopy and imaging~\cite{Maussang05,Wang10}, electron impact ionization in combination with depletion techniques for state selectivity~\cite{Motsch07,Gloeckner15a,Wu16}, and ionization with a femtosecond laser~\cite{Meng15}. For experiments with Feshbach associated alkali dimers, molecules are typically dissociated back into atoms for their detection~\cite{Moses16}. The most widely used of these techniques, REMPI and LIF, rely on favorable properties of the molecules: sufficiently long-lived electronically excited states in the case of REMPI, and electronically excited states which primarily decay radiatively in the case of LIF. REMPI and femtosecond ionization only detect molecules in the focus of a high-power laser. Except for LIF or absorption measurements applied to the relatively unique molecule species with highly diagonal Franck-Condon factors~\cite{DiRosa04}, all these techniques destroy the molecules in the process of detection. A generally applicable technique to detect molecules over a large volume with high efficiency, preferably nondestructively, would thus be of immense value.

In this paper, we propose detecting polar molecules via resonant energy transfer between molecules and Rydberg atoms. Such energy transfer processes have been studied in detail for Rydberg-Rydberg interactions, and have even been observed for molecule-Rydberg interactions, but not in the context of detecting molecules~\cite{Ravets14,Smith78}. We calculate the interaction cross section for the energy transfer process, demonstrating that huge values of more than $10^{-6}\,$cm$^2$ are possible for suitably chosen experimental parameters. Experimental scenarios for detecting molecular beams or trapped molecule ensembles are discussed, demonstrating that a near unit efficiency detection of molecules is possible, applicable to basically any polar molecule species.

We consider the dipole-dipole interaction energy between a Rydberg atom and a polar molecule. For a molecule dipole moment $d_{\rm mol}$ and a Rydberg dipole moment $d_{\rm Ryd}$ oriented in parallel, separated by a distance $r$, with an angle $\theta$ between the orientation of the dipoles and the direction of the interparticle separation, the dipole-dipole energy is given by
\begin{equation}\label{Edd}
E_{d,d}=\frac{d_{\rm mol}\,d_{\rm Ryd}(1-3\cos^2\theta)}{4\pi\epsilon_0r^3}.
\end{equation}
For a molecule dipole moment of $d_{\rm mol}=1$\,Debye and a Rydberg dipole moment of $d_{\rm Ryd}=6600$\,Debye, separated by $r=1\,\mu$m and oriented side by side with $\theta=\pi/2$, we obtain $E_{d,d}=1$\,MHz$\times\,h$, where $h$ is Planck's constant. This relatively large interaction energy for relatively large interparticle separation makes detection of polar molecules via Rydberg atoms favorable.

A relatively straightforward approach to detect polar molecules with Rydberg atoms would be to bring an atom and a molecule into close proximity and to detect the energy shift given by Eq.~\ref{Edd} on a Rydberg transition. This idea has in fact been suggested previously in a paper by Kuznetsova {\it et al.}~\cite{Kuznetsova11} to read out molecular qubits for quantum information processing. However, this detection method is probably only possible in a highly controlled environment, with precise control over the molecule-Rydberg-atom separation and no other Rydberg atoms in the vicinity which would cause much larger energy shifts.

\begin{figure}[t]
\centering
\includegraphics[width=0.45\textwidth]{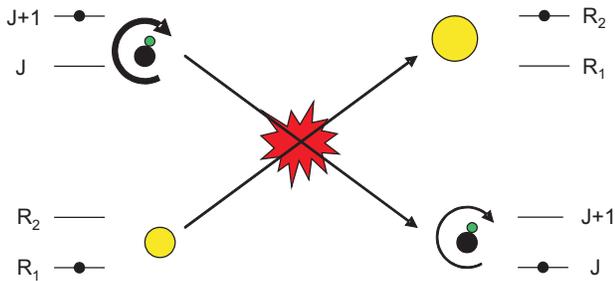}
\caption{(Color online). F\"orster resonance energy transfer between a molecule and a Rydberg atom, as discussed in the main text.}\label{collision}
\end{figure}

A much more robust approach to detecting polar molecules with Rydberg atoms is to make use of F\"orster resonance energy transfer, as shown in Fig.~\ref{collision}. The key ingredients are a pair of molecular states $J$ and $J+1$ and a pair of Rydberg states $R_1$ and $R_2$ with a large dipole transition moment as well as an equal energy separation between both pairs of states. We use $J$ and $J+1$ to denote the molecular states in reference to a pair of neighboring rotational states although any pair of molecular states can be considered. In this case, for a molecule initially in state $J+1$ and a Rydberg atom initially in state $R_1$, F\"orster resonance energy transfer can result in the molecule ending in state $J$ and the Rydberg atom ending in state $R_2$ after the two particles fly past one another. For reversed initial conditions, the reverse process is of course equally possible. State sensitive detection of the Rydberg atom, for example via state sensitive field ionization~\cite{Gallagher77,Jeys80,Guertler04}, thus provides a signal which depends on the presence of molecules.

We calculate the energy transfer cross section for a molecule-Rydberg-atom collision. We consider a classical trajectory, with the molecule-Rydberg-atom separation versus time $t$ given by
\begin{equation}
\mathbf{r}(t)=b\,\mathbf{\hat{x}}+v\,t\,\mathbf{\hat{z}}.
\end{equation}
Here, $v$ is the relative velocity and $b$ is the impact parameter. The relevant internal states for the molecule and Rydberg atom are $|J+1,R_1\rangle$ and $|J,R_2\rangle$, with the system initially in the state $|J+1,R_1\rangle$. The Hamiltonian for the internal states is given by
\begin{equation}\label{Hamiltonian}
\hat{H}=\begin{pmatrix} 0 & E_{d,d}\\ E_{d,d} & \Delta\end{pmatrix},
\end{equation}
where $E_{d,d}$ is the dipole-dipole interaction energy, and $\Delta$ is the energy mismatch between the pairs of states.

For simplicity, we only consider $\Delta M=0$ transitions such that the transition dipole moment of both the molecule and the Rydberg atom are oriented along an externally applied electric field. In this case, the dipole-dipole interaction energy is given by Eq.~\ref{Edd}, with
\begin{equation}\label{dipole mol}
d_{\rm mol}=\langle J|\mathbf{d}\cdot\mathbf{\hat{n}}|J+1\rangle,
\end{equation}
\begin{equation}\label{dipole Ryd}
d_{\rm Ryd}=\langle R_1|\mathbf{d}\cdot\mathbf{\hat{n}}|R_2\rangle,
\end{equation}
and
\begin{equation}
\cos(\theta)=\frac{\mathbf{\hat{n}}\cdot\mathbf{r}}{|\mathbf{r}|}.
\end{equation}
Here, $\mathbf{\hat{n}}=(n_x,n_y,n_z)$ is a unit vector along the direction of the externally applied electric field. Note that the dipole moments in Eqs.~\ref{dipole mol} and \ref{dipole Ryd} are transition dipole moments rather than static dipole moments. Thus, no external electric field is required to orient the molecule or Rydberg atom, and the electric field is only used to define a quantization axis and, as discussed below, to match the molecule and Rydberg atom transition frequencies.

We assume zero energy mismatch between the pairs of states $J$, $J+1$ and $R_1$, $R_2$, i.e.\ $\Delta=0$, in which case the time evolution for the Hamiltonian in Eq.~\ref{Hamiltonian} can be easily solved to obtain
\begin{equation}
|\Psi(t)\rangle=\cos(\Phi(t))|J+1,R_1\rangle-i\sin(\Phi(t))|J,R_2\rangle,
\end{equation}
with
\begin{equation}
\Phi(t)=\int_{-\infty}^tE_{d,d}(t')dt'/\hbar.
\end{equation}
This integral can be solved analytically, and the probability $p_{1\rightarrow2}$ to end in state $|J,R_2\rangle$ is
\begin{equation}
p_{1\rightarrow2}=\sin^2(\Phi(\infty))=\sin^2\left(\frac{2d_{\rm mol}\,d_{\rm Ryd}(n_y^2-n_x^2)}{4\pi\epsilon_0\,b^2\,v\,\hbar}\right).
\end{equation}
Integrating this probability over all possible values and orientations of $b$ in the $x-y$ plane, we obtain the energy transfer cross section
\begin{equation}\label{crosssection}
\sigma=\frac{2\pi d_{\rm mol}\,d_{\rm Ryd}(1-n_z^2)}{4\pi\epsilon_0\,v\,\hbar}.
\end{equation}

The validity of Eq.~\ref{crosssection} depends on two conditions being fulfilled. First, Eq.~\ref{Edd} for the dipole-dipole interaction energy is only valid when the molecule-Rydberg separation is sufficiently larger than the radius $r_{\rm Ryd}$ of the Rydberg atom. Thus, Eq.~\ref{crosssection} is only valid for $\sigma\gg r_{\rm Ryd}^2$. Second, Eq.~\ref{crosssection} requires the energy mismatch $\Delta$ between the two molecular states and the two Rydberg states to be sufficiently small. From a rough back-of-the-envelope estimate, the maximum value of $\Delta$ is given by
\begin{equation}\label{Deltamax}
\Delta_{\rm max}\approx\sqrt{\frac{4\pi\epsilon_0}{d_{\rm mol}\,d_{\rm Ryd}}}(\hbar\,v)^{3/2},
\end{equation}
based on $\Delta$ needing to be smaller than the dipole-dipole interaction energy at a molecule-Rydberg atom separation of $r\approx\sqrt{\sigma}$. The Rydberg transition can be tuned close to the molecule transition using external electric fields.

\begin{table*}
\centering
\begin{tabular}{l|ccc|c|c|c}
Molecule:&&H$_2$CO&&CH$_3$F&RbCs&LiCs\\
\hline
Molecule&&$(1,1,1,1)\leftrightarrow$&&$(0,0,0)\leftrightarrow$&$(0,0)\leftrightarrow (1,0)$&$(1,0)\leftrightarrow (2,0)$\\
transition&&$(1,1,0,1)$&&$(1,0,0)$&&\\
(notation):&&$(J,K_A,K_C,M)$&&$(J,K,M)$&$(J,M)$&$(J,M)$\\
$f_{\rm mol}$:&&4.83\,GHz&&51.1GHz&1.02\,GHz&23.3\,Ghz\\
$d_{\rm mol}$:&&1.16\,Debye&&1.07\,Debye&0.72\,Debye&2.85\,Debye\\
Rubidium&$90S_{1/2}\leftrightarrow90P_{3/2}$,&$76P_{3/2}\leftrightarrow75D_{5/2}$,&$78D_{5/2}\leftrightarrow77F_{7/2}$&$36P_{3/2}\leftrightarrow35D_{3/2}$&$149S_{1/2}\leftrightarrow149P_{3/2}$&$47D_{5/2}\leftrightarrow46F_{7/2}$\\
transition:&&&&&&\\
$f_{\rm Ryd}$:&4.87\,GHz&4.89\,GHz&4.79\,GHz&51.8\,GHz&1.03\,GHz&22.6\,GHz\\
$d_{\rm Ryd}$:&10240\,Debye&9165\,Debye&9891\,Debye&1917\,Debye&28760\,Debye&3541\,Debye\\
$\sigma$ ($v=1$\,m/s):&$7.1\times10^{-7}$\,cm$^2$&$6.3\times10^{-7}$\,cm$^2$&$6.8\times10^{-7}$\,cm$^2$&$1.2\times10^{-7}$\,cm$^2$&$1.23\times10^{-6}$\,cm$^2$&$6.0\times10^{-7}$\,cm$^2$\\
$\tau (300\,$K):&260\,$\mu$s, 310\,$\mu$s&220\,$\mu$s, 170\,$\mu$s&190\,$\mu$s, -&38\,$\mu$s, 26\,$\mu$s&790\,$\mu$s, 880\,$\mu$s&56\,$\mu$s, -\\
$\Gamma_{\rm BB}$ ($300\,$K):&181\,Hz&146\,Hz&163\,Hz&717\,Hz&64\,Hz&466\,Hz\\
$\chi$:&$0.67\%$&$1.3\%$&$1.2\%$&$5.1\%$&$0.16\%$&$8.4\%$
\end{tabular}
\caption{Possible experimental parameters for the detection of four different molecule species in the states indicated via Rubidium Rydberg atoms. Various Rydberg transitions are possible for each molecule transition, as shown explicitly for formaldehyde. For the other molecules only a single Rydberg transition is considered. We ignore the hyperfine structure for RbCs and LiCs. For H$_2$CO and CH$_3$F the hyperfine structure is negligible. The Rydberg transition frequencies and transition dipole moments are calculated using quantum defect theory~\cite{Li03,Bhatti81,Marinescu94}. The Rydberg transition dipole moment is for $\Delta M=0$ transitions between the $|M|=1/2$ states. $\tau$ is the lifetime of the  nS, nP, and nD Rydberg states, calculated according to Ref.~\cite{Beterov09}. $\Gamma_{\rm BB}$ is the blackbody induced transition rate for the Rydberg transition, calculated from $f_{\rm Ryd}$ and $d_{\rm Ryd}$. $\chi$ is the single shot detection efficiency as defined in the main text.}\label{table}
\end{table*}

Possible values for the interaction cross section $\sigma$ are considered in table~\ref{table} for four molecule species which are relevant to contemporary experiments~\cite{Prehn16,Meng15,Zeppenfeld12,Takekoshi14,Kraft06}. The interaction cross section is huge, particularly for low collision energies. Various angular momentum states can be used for the Rydberg transitions for each molecule transition. For Rubidium Rydberg atoms, the s-p transitions have the highest transition dipole moments. However, the lower principle quantum numbers for the p-d and d-f transitions is advantageous for creating a high-density Rydberg gas as discussed below. Note that the transition dipole moment of the strongest Rydberg transition close to a given molecule frequency $f_{\rm mol}$ scales as $d_{\rm Ryd}\propto f_{\rm mol}^{-2/3}$, allowing the interaction cross-section for other molecule transitions to be estimated.

Probably the most important experimental issue for implementing detection of molecules via Rydberg atoms is the density of Rydberg atoms that can be maintained for a given amount of time. Thus, at high Rydberg densities, Rydberg-Rydberg interactions cause redistribution of the Rydberg population among different states and even avalanche ionization~\cite{Li06}. In Ref.~\cite{Li06}, it is found that this occurs at a rate of about $300$\,kHz for Rydberg atoms in the 46D state at a density of $10^9\,$cm$^{-3}$. This rate roughly scales with the Rydberg principle quantum number $n$ as $n^4$, but depends on details such as the attractive or repulsive nature of the long-range van-der-Waals interaction between the Rydberg atoms~\cite{Li06}. For the examples below, we assume that a Rydberg atom density of at most $10^9\,$cm$^{-3}$ can be maintained for $1\,\mu$s for $n=50$, with the density inversely proportional to the duration and scaling as $n^{-4}$ with $n$.

Interestingly, since the achievable Rydberg atom density scaling as $n^{-4}$ dominates over the Rydberg transition dipole moment scaling as $n^2$, Rydberg based detection of molecules will likely work best for large molecule transition frequencies and correspondingly low $n$ where the interaction cross section is smaller. In this sense the examples of H$_2$CO and RbCs in table~\ref{table} are chosen poorly, since they focus on lower molecular transition frequencies to exemplify the large interaction cross sections that can be obtained. In principle, molecule transitions with much higher transition frequencies exist, but even for the transitions in table~\ref{table} large detection efficiencies are possible for achievable Rydberg densities, as discussed next.

As a first experimental scenario for detecting molecules via Rydberg atoms, we consider detection of molecules in a beam traveling at a velocity $v\lesssim100\,$m/s. This is very roughly the limit on $v$ such that the condition $\sigma\gg r_{\rm Ryd}^2$ discussed above holds. Such a beam might be generated, for example, by Stark deceleration~\cite{Bethlem02}, buffergas cooling~\cite{Maxwell05,Wu16}, or velocity filtering~\cite{Sommer10}. We consider Rydberg atoms spread over a length $L$ along the beam and assume that a Rydberg density $\rho_{\rm Ryd}$ is maintained for a time $T=L/v$, corresponding to the time for a molecule to traverse the Rydberg cloud. $T$ might reasonably be in the range $1-100\,\mu$s. According to the previous discussion, the maximum possible Rydberg density is $(n/50)^{-4}/T\times10^9\,\mu$s\,cm$^{-3}$. For a molecule-Rydberg cross section $\sigma$, the probability $\chi$ for a molecule traversing the Rydberg cloud to interact with an atom is
\begin{equation}\label{singleshot}
\chi=L\,\sigma\,\rho_{\rm Ryd}=v\,\sigma\,(n/50)^{-4} 10^9\,\mu{\rm s\,cm}^{-3}.
\end{equation}
Since $\sigma\propto1/v$, $\chi$ is independent of the molecule velocity and the interaction length $L$, and is tabulated for the transitions considered in table~\ref{table}. $\chi$ is the maximum probability to detect a molecule when producing Rydberg atoms in a single shot, and is as large as $8\,\%$ for the transitions in table~\ref{table}. Since Rydberg atoms can be created over a large area of $1\,$mm$^2$ or more without affecting the detection efficiency, molecules can be detected over a large area which is a key advantage of this method.

Two issues affecting detection of beams of molecules are the molecule pulse length and signal to noise ratio. For a molecule pulse duration shorter than the Rydberg state lifetime, Rydberg atoms can be produced in a single shot. This is often the case for Stark deceleration, with a pulse length of, e.g., $3\,$mm independent of velocity~\cite{Bethlem02}. For buffergas cooling and velocity filtering, molecule pulses are substantially longer or even continuous. In this case Rydberg atoms will need to be produced repeatedly to obtain a high detection efficiency.

Molecule detection will work best if molecules induce transitions to the final Rydberg state at a faster rate than background processes. The former rate is given by $v\,\sigma\,\rho_{\rm mol}$, where $\rho_{\rm mol}$ is the molecule density. We consider two background processes, Rydberg-Rydberg interactions and blackbody radiation. While the exact effect of Rydberg-Rydberg interactions would need to be investigated, the results in Ref.~\cite{Li06} indicate that most Rydberg atoms remain in the initial state even when a substantial fraction of Rydberg atoms has been ionized due to Rydberg-Rydberg interactions. The effect of Rydberg-Rydberg interactions can be suppressed by reducing the Rydberg density and correspondingly increasing the interaction time to maintain the same detection efficiency.

Unlike Rydberg-Rydberg interactions, the rate of blackbody induced transitions is constant at a given temperature and can be directly compared to the transition rate due to molecules. For the transitions in table~\ref{table}, the blackbody transition rate and the molecule induced transition rate are equal for molecule densities of $\sim2.4\times10^6\,$cm$^{-3}$, $6.0\times10^7\,$cm$^{-3}$, $5.2\times10^5\,$cm$^{-3}$, and $7.8\times10^6\,$cm$^{-3}$ for H$_2$CO, CH$_3$F, RbCs, and LiCs, respectively. Lower densities can of course be detected, but in this case the majority of signal would be due to blackbody radiation and would need to be subtracted. Alternatively, a low-background detection could be performed in a cryogenic environment.

As a second experimental scenario, we consider molecules and Rydberg atoms in a trap. The derivation of Eq.~\ref{singleshot} applies almost identically to trapped molecules, leading to the same single shot detection efficiency. A key advantage of detecting molecules in a trap is that Rydberg atoms can be excited repeatedly in the same volume to probe the same molecules, allowing the detection efficiency to be increased arbitrarily close to unity.  After several detection events, a substantial fraction of molecules will have been transferred to the second molecule state, reducing the signal. At that point, atoms can be alternatingly produced in the two Rydberg states, thereby shuffling the molecules back and forth between the two molecule states and allowing a molecule to be detected multiple times. In this way, even a sub-shot-noise detection of the molecule number is possible.

A key consideration for detecting trapped molecules is the effect of the trapping fields. First, strong electric or microwave fields to trap molecules will field ionize the Rydberg atoms, and the trapping fields would thus need to be switched off during detection. Optical or magnetic traps might thus be preferable. Second, the trapping fields will cause inhomogeneous broadening of the molecule and Rydberg transitions. According to Eq.~\ref{Deltamax}, for a velocity of $1\,$m/s, the energy mismatch between the molecule and Rydberg transitions must be less than about $50\,$kHz$\times\,h$ for the transitions in table~\ref{table}. This is roughly three orders of magnitude less than the kinetic energy of a molecule at $1\,$m/s. Thus, either the trapping fields will need to be switched off during detection independent of the effect on the Rydberg atoms, pairs of states with almost identical energy shift vs. trapping field will be needed, or a decrease in the energy transfer cross section will need to be tolerated.

As a specific example for detecting trapped molecules, we consider an ensemble of molecules at roughly $1\,$mK occupying a volume of $1\,$mm$^3$ in a quadrupole electric trap, superimposed on a magneto-optical trap for atoms. A single detection sequence might consist of switching off the electric trap ($\sim1\,\mu$s), exciting Rydberg atoms with a laser ($\sim1\,\mu$s), interrogating for $10\,\mu$s, field-ionizing the Rydberg atoms ($\sim1\,\mu$s), and switching the electric trap back on ($\sim1\,\mu$s). The previous discussion on background processes applies equally to trapped molecules, and at least 500 to 60000 trapped molecules would be needed so that molecule induced transitions dominate over blackbody radiation induced transitions.

Slightly modifying the previous example corresponds to detection of a single molecule. Thus, a single molecule and a single Rydberg atom confined to a volume of $(10\,\mu$m$)^3$ is equivalent to a density of $10^9$\,cm$^{-3}$. For a cross section of $10^{-6}\,$cm$^2$ at $1$\,m/s, a molecule-Rydberg-atom interaction will occur on average every $10\,\mu$s.

The proposed molecule detection technique is likely to be of great benefit for a wide variety of experiments. Thus, we have shown that Rydberg detection of molecules can be a highly efficient technique to detect molecular beams or trapped ensembles of molecules. Rydberg detection would be of particular benefit for the many molecule species without suitable electronic states for REMPI- or LIF-based detection schemes, as its only requirement is a molecule species with a permanent electric dipole moment in the molecule frame. Rydberg detection can be implemented with minimal overhead in current experiments with alkali dimers as well as future experiments using ultracold alkali atoms for sympathetic cooling of molecules, as such experiments automatically contain alkali atoms which could be excited to Rydberg states for molecule detection. As a final note, we emphasize that Rydberg detection is nondestructive, in that the molecules persist in a well-defined state after detection. This would be particularly beneficial, e.g., for quantum information processing with arrays of molecules where the positions of individual molecules need to be determined before an experiment is performed.


\begin{thebibliography}{10}

\bibitem{Shalm15}
L.K.~Shalm {\it et al.},
"Strong Loophole-Free Test of Local Realism."
Phys. Rev. Lett. {\bf 115}, 250402 (2015).

\bibitem{Giustina15}
M. Giustina {\it et al.},
"Significant-Loophole-Free Test of Bell's Theorem with Entangled Photons."
Phys. Rev. Lett. {\bf 115}, 250401 (2015).

\bibitem{Bakr09}
W.S.~Bakr, J.I.~Gillen, A.~Peng, S.~F\"olling, and M.~Greiner,
"A quantum gas microscope for detecting single atoms in a Hubbard-regime optical lattice."
Nature {\bf 462}, 74--77 (2009).

\bibitem{Endres11}
M.~Endres, M.~Cheneau, T.~Fukuhara, C.~Weitenberg, P.~Schau\ss, C.~Gross, L.~Mazza, M.C.~Ba\~nuls, L.~Pollet, I.~Bloch, and S.~Kuhr,
"Observation of Correlated Particle-Hole Pairs and String Order in Low-Dimensional Mott Insulators."
Science {\bf 334}, 200--203 (2011).

\bibitem{Cheneau12}
M.~Cheneau, P.~Barmettler, D.~Poletti, M.~Endres, P.~Schau\ss, T.~Fukuhara, C.~Gross, I.~Bloch, C.~Kollath, and S.~Kuhr,
"Light-cone-like spreading of correlations in a quantum many-body system."
Nature {\bf 481}, 484--487 (2012).

\bibitem{DeMille02}
D.~DeMille,
"Quantum Computation with Trapped Polar Molecules."
Phys. Rev. Lett. {\bf 88}, 067901 (2002).

\bibitem{DeMille08}
D.~DeMille, S.B.~Cahn, D.~Murphree, D.A.~Rahmlow, and M.G. Kozlov,
"Using Molecules to Measure Nuclear Spin-Dependent Parity Violation."
Phys. Rev. Lett. {\bf 100}, 023003 (2008).

\bibitem{Baranov12}
M.A.~Baranov, M.~Dalmonte, G.~Pupillo, and P.~Zoller,
"Condensed matter theory of dipolar quantum gases."
Chem. Rev. {\bf 112}, 5012 (2012).

\bibitem{Gloeckner15b}
R.~Gl\"ockner, A.~Prehn, B.G.U.~Englert, G.~Rempe, and M.~Zeppenfeld,
"Rotational Cooling of Trapped Polyatomic Molecules."
Phys. Rev. Lett. {\bf 115}, 233001 (2015).

\bibitem{Prehn16}
A.~Prehn, M.~Ibr\"ugger, R.~Gl\"ockner, G.~Rempe, and M.~Zeppenfeld,
"Optoelectrical Cooling of Polar Molecules to Submillikelvin Temperatures."
Phys. Rev. Lett. {\bf 116}, 063005 (2016).

\bibitem{Norrgard16}
E.B.~Norrgard, D.J.~McCarron, M.H.~Steinecker, M.R.~Tarbutt, and D.~DeMille,
"Submillikelvin Dipolar Molecules in a Radio-Frequency Magneto-Optical Trap."
Phys. Rev. Lett. {\bf 116}, 063004 (2016).

\bibitem{Moses16}
S.A.~Moses, J.P.~Covey, M.T.~Miecnikowski, B.~Yan, B.~Gadway, J.~Ye, and D.S.~Jin,
"Creation of a low-entropy quantum gas of polar molecules in an optical lattice."
Science {\bf 350}, 659--662 (2016).

\bibitem{Bethlem00}
H.L.~Bethlem, G.~Berden, F.M.H.~Crompvoets, R.T.~Jongma, A.J.A.~van~Roij, and G. Meijer,
"Electrostatic trapping of ammonia molecules."
Nature {\bf 406}, 491--494 (2000).

\bibitem{Bertsche10}
B.~Bertsche and A.~Osterwalder,
"State-selective detection of velocity-filtered ND$_3$ molecules."
Phys. Rev. A {\bf 82}, 033418 (2010).

\bibitem{Weinstein98}
J.D.~Weinstein, R.~DeCarvalho, T.~Guillet, B.~Friedrich, and J.M.~Doyle,
"Magnetic trapping of calcium monohydride molecules at millikelvin temperatures."
Nature {\bf 395}, 148--150 (1998).

\bibitem{Shuman09}
E.S.~Shuman, J.F.~Barry, D.R.~Glenn, and D. DeMille,
"Radiative Force from Optical Cycling on a Diatomic Molecule."
Phys. Rev. Lett. {\bf 103}, 223001 (2009).

\bibitem{Maussang05}
K.~Maussang, D.~Egorov, J.S.~Helton, S.V.~Nguyen, and J.M.~Doyle,
"Zeeman Relaxation of CaF in Low-Temperature Collisions with Helium."
Phys. Rev. Lett. {\bf 94}, 123002 (2005).

\bibitem{Wang10}
D.~Wang, B.~Neyenhuis, M.H.G.~de~Miranda, K.-K.~Ni, S.~Ospelkaus, D.S.~Jin, and J.~Ye,
"Direct absorption imaging of ultracold polar molecules."
Phys. Rev. A. {\bf 81}, 061404(R) (2010).

\bibitem{Motsch07}
M.~Motsch, M.~Schenk, L.D.~van Buuren, M.~Zeppenfeld, P.W.H.~Pinkse, and G.~Rempe,
"Internal-state thermometry by depletion spectroscopy in a cold guided beam of formaldehyde."
Phys. Rev. A {\bf 76}, 061402(R) (2007).

\bibitem{Gloeckner15a}
R.~Gl\"ockner, A.~Prehn, G.~Rempe, and M.~Zeppenfeld,
"Rotational state detection of electrically trapped polyatomic molecules."
New J. Phys. {\bf 17}, 055022 (2015).

\bibitem{Wu16}
X.~Wu, T.~Gantner, M.~Zeppenfeld, S.~Chervenkov, and G.~Rempe,
"Thermometry of Guided Molecular Beams from a Cryogenic Buffer-Gas Cell."
ChemPhysChem {\bf 17}, 3631--3640 (2016).

\bibitem{Meng15}
C.~Meng, A.P.P.~van~der~Poel, C.~Cheng, and H.L.~Bethlem,
"Femtosecond laser detection of Stark-decelerated and trapped methylfluoride molecules."
Phys. Rev. A {\bf 92}, 023404 (2015).

\bibitem{DiRosa04}
M.D.~Di~Rosa,
"Laser-cooling molecules."
Eur. Phys. J. D {\bf 31}, 395--402 (2004).

\bibitem{Ravets14}
S.~Ravets, H.~Labuhn, D.~Barredo, L.~B\'eguin, T.~Lahaye, and A.~Browaeys,
"Coherent dipole-dipole coupling between two single Rydberg atoms at an electrically-tuned F\"orster resonance."
Nat. Phys. {\bf 10}, 914--917 (2014).

\bibitem{Smith78}
K.A.~Smith, F.G.~Kellert, R.D.~Rundel, F.B.~Dunning, and R.F.~Stebbings,
"Discrete Energy Transfer in Collisions of Xe(nf) Rydberg Atoms with NH$_3$ Molecules."
Phys. Rev. Lett. {\bf 40}, 1362 (1978).

\bibitem{Kuznetsova11}
E.~Kuznetsova, S.T.~Rittenhouse, H.R.~Sadeghpour, and S.F.~Yelin,
"Rydberg atom mediated polar molecule interactions: a tool for molecular-state conditional quantum gates and individual addressability."
Phys. Chem. Chem. Phys. {\bf 13}, 17115 (2011).

\bibitem{Gallagher77}
T.F.~Gallagher, L.M.~Humphrey, W.E.~Cooke, R.M.~Hill, and S.A.~Edelstein,
"Field ionization of highly excited states of sodium."
Phys. Rev. A {\bf 16}, 1098 (1977).

\bibitem{Jeys80}
T.H.~Jeys, G.W.~Foltz, K.A.~Smith, E.J.~Beiting, F.G.~Kellert, F.B.~Dunning, and R.F.~Stebbings,
"Diabatic Field Ionization of Highly Excited Sodium Atoms."
Phys. Rev. Lett. {\bf 44}, 390 (1980).

\bibitem{Guertler04}
A.~G\"urtler and W.J.~van~der~Zande,
"l-state selective field ionization of rubidium Rydberg states."
Phys. Lett. A {\bf 324}, 315--320 (2004).

\bibitem{Zeppenfeld12}
M.~Zeppenfeld, B.G.U.~Englert, R.~Gl\"ockner, A.~Prehn, M.~Mielenz, C.~Sommer, L.D.~van~Buuren, M.~Motsch, and G.~Rempe,
"Sisyphus cooling of electrically trapped polyatomic molecules."
Nature {\bf 491}, 570--573 (2012).

\bibitem{Takekoshi14}
T.~Takekoshi, L.~Reichs\"ollner, A.~Schindewolf, J.M.~Hutson, C.R.~Le Sueur, O.~Dulieu, F.~Ferlaino, R.~Grimm, and H.-C.~N\"agerl,
"Ultracold Dense Samples of Dipolar RbCs Molecules in the Rovibrational and Hyperfine Ground State."
Phys. Rev. Lett. {\bf 113}, 205301 (2014).

\bibitem{Kraft06}
S.D.~Kraft, P.~Staanum, J.~Lange, L.~Vogel, R.~Wester, and M.~Weidem\"uller,
"Formation of ultracold LiCs molecules."
J. Phys. B {\bf 39}, S993 (2006).

\bibitem{Li03}
W.~Li, I.~Mourachko, M.W.~Noel, and T.F.~Gallagher,
"Millimeter-wave spectroscopy of cold Rb Rydberg atoms in a magneto-optical trap: Quantum defects of the $ns$, $np$, and $nd$ series."
Phys. Rev. A {\bf 67}, 052502 (2003).

\bibitem{Bhatti81}
S.A.~Bhatti, C.L.~Cromer, and W.E.~Cooke,
"Analysis of the Rydberg character of the $5d7d^1D_2$ state of barium."
Phys. Rev. A {\bf 24}, 161 (1981).

\bibitem{Marinescu94}
M.~Marinescu, H.R.~Sadeghpour, and A.~Dalgarno,
"Dispersion coefficients for alkali-metal dimers."
Phys. Rev. A {\bf 49}, 982 (1994).

\bibitem{Beterov09}
I.I.~Beterov, I.I.~Ryabtsev, D.B.~Tretyakov, and V.M.~Entin,
"Quasiclassical calculations of blackbody-radiation-induced depopulation rates and effective lifetimes of Rydberg $nS$, $nP$, and $nD$ alkali-metal atoms with $n\le80$."
Phys. Rev. A {\bf 79}, 052504 (2009).

\bibitem{Li06}
W.~Li, P.J.~Tanner, Y.~Jamil, and T.F.~Gallagher,
"Ionization and plasma formation in high n cold Rydberg samples."
Eur. Phys. J. D {\bf 40}, 27--35 (2006).

\bibitem{Bethlem02}
H.L.~Bethlem, F.M.H.~Crompvoets, R.T.~Jongma, S.Y.T.~van~de~Meerakker, and G.~Meijer,
"Deceleration and trapping of ammonia using time-varying electric fields."
Phys. Rev. A {\bf 65}, 053416 (2002).

\bibitem{Maxwell05}
S.E.~Maxwell, N.~Brahms, R.~deCarvalho, D.R.~Glenn, J.S.~Helton, S.V.~Nguyen, D.~Patterson, J.~Petricka, D.~DeMille, and J.M.~Doyle,
"High-Flux Beam Source for Cold, Slow Atoms or Molecules."
Phys. Rev. Lett. {\bf 95}, 173201 (2005).

\bibitem{Sommer10}
C.~Sommer, M.~Motsch, S.~Chervenkov, L.D.~van~Buuren, M.~Zeppenfeld, P.W.H.~Pinkse, and G.~Rempe,
"Velocity-selected molecular pulses produced by an electric guide."
Phys. Rev. A {\bf 82}, 013410 (2010).

\end{thebibliography}

\begin{acknowledgments}
Many thanks to Daniel Tiarks, Stephan D\"urr, Peter Schau\ss, Gerhard Rempe, and Daniel Comparat for helpful discussions.
\end{acknowledgments}

\bibliographystyle{unsrt}

\end{document}